\documentclass[english,prb,twocolumn,superscriptaddress,amsmath,amssymb,a4paper]{revtex4}
\usepackage[T1]{fontenc}
\usepackage[latin1]{inputenc}
\usepackage{graphicx}
\usepackage[top=0.85in,bottom=0.9in,,left=0.75in,right=0.75in]{geometry}
\usepackage{booktabs}
\usepackage{color}

\makeatletter
\usepackage{dcolumn}
\usepackage{bm}
\usepackage{epsfig}

\newcommand{\ignore}[1]{}

\usepackage{babel}
\makeatother

\usepackage{sidecap}
\sidecaptionvpos{figure}{c}

\begin{document}

\title{Electronic and Magnetic Structure of Infinite-layer $\textrm{NdNiO}_2$:\\
Trace of Antiferromagnetic Metal}

\author{Zhao Liu}
\affiliation{Hefei National Laboratory for Physical Sciences at the Microscale,
CAS Key Laboratory of Strongly-Coupled Quantum Matter Physics,
University of Science and Technology of China, Hefei, Anhui 230026, China}

\author{Zhi Ren}
\affiliation{Institute of Natural Sciences, Westlake Institution of Advanced Study and
School of Science, Westlake University, Hangzhou 300024, China}

\author{W. Zhu} %\thanks{E-mail: zhuwei@westlake.edu.cn}
\affiliation{Institute of Natural Sciences, Westlake Institution of Advanced Study and
School of Science, Westlake University, Hangzhou 300024, China}

\author{Z. F. Wang} 
\affiliation{Hefei National Laboratory for Physical Sciences at the Microscale,
CAS Key Laboratory of Strongly-Coupled Quantum Matter Physics,
University of Science and Technology of China, Hefei, Anhui 230026, China}

\author{Jinlong Yang} \thanks{E-mail: jlyang@ustc.edu.cn}
\affiliation{Hefei National Laboratory for Physical Sciences at the Microscale,
Synergetic Innovation Center of Quantum Information and Quantum Physics,
University of Science and Technology of China, Hefei, Anhui 230026, China}

%\date{\today}% It is always \today, today, % but any date may be explicitly specified

\begin{abstract}
The recent discovery of Sr-doped infinite-layer nickelate ${\rm NdNiO_2}$
[D. Li \textit{et al.} Nature \textbf{572}, 624 (2019)] offers an exciting platform for investigating
unconventional superconductivity in nickelate-based compounds. In this work, we present a first-principles calculations for
the electronic and magnetic properties of undoped parent ${\rm NdNiO_2}$.
Intriguingly, we found that: 1) the paramagnetic phase has complex Fermi pockets with 3D characters near the Fermi level;
2) by including electron-electron interactions, $3d$-electrons of Ni tend to form $(\pi, \pi, \pi)$ antiferromagnetic ordering at low temperatures;
3) with moderate interaction strength, $5d$-electrons of Nd contribute small Fermi pockets that could weaken the magnetic order
akin to the self-doping effect. Our results provide a plausible interpretation for the experimentally observed resistivity minimum
and Hall coefficient drop. Moreover, we elucidate that antiferromagnetic ordering in ${\rm NdNiO_2}$ is relatively weak,
arising from the small exchange coupling between $3d$-electrons of Ni and also hybridization with $5d$-electrons of Nd.
\end{abstract}

\maketitle

%\section{I. Introduction}
Since the discovery of high-temperature (high-$T_c$) superconductivity in cuprates \cite{htCu},
extensive effort has been devoted to investigate unconventional superconductors, ranging from
non-oxide compounds \cite{MgB, HS} to iron-based materials \cite{htFe1, htFe2}. Exploring high-$T_c$
materials could provide a new platform to understand the fundamental physics behind high-$T_c$ phenomenon,
thus is quite valuable. Very recently, the discovery of superconductivity in Sr-doped ${\rm NdNiO_2}$ \cite{Nat1}
potentially raises the possibility to realize high-$T_c$ in nickelate family \cite{pre1, pre2}.

One key experimental observation for the infinite-layer ${\rm NdNiO_2}$ is that its resistivity exhibits a minimum
around 70 K and an upturn at a lower temperature \cite{Nat1}. At the same time that the resistivity reaches minimum,
the Hall coefficient drops towards a large value, signalling the loss of charge carriers \cite{Nat1}. Interestingly,
no long-range magnetic order has been observed in powder neutron diffraction on ${\rm NdNiO_2}$ when temperature is
down to 1.7 K \cite{Nat1}. This greatly challenges the existing theories, since it is generally believed that
magnetism holds the key to understand unconventional superconductivity \cite{Cu1, Cu2, Fe1, Fe2}. Therefore,
it is highly desirable to study the magnetic properties of undoped parent ${\rm NdNiO_2}$ and elucidate its
experimental indications.

In this work, the electronic and magnetic properties of ${\rm NdNiO_2}$ are systemically studied by first-principles
calculations combined with classical Monte Carlo calculations. Firstly, the paramagnetic (PM) phase is studied. Its Fermi surface includes one large sheet and two electron pockets
at $\Gamma$ and A point, respectively. This can be described by a three-band low-energy effective model that captures the main physics of exchange coupling mechanism. Then, the magnetic properties are studied by including Hubbard U and $(\pi, \pi, \pi)$
anti-ferromagnetic (AFM) ordering is confirmed to be the magnetic ground state. Most significantly, the Fermi surface of
AFM phase is simpler than that of PM phase, demonstrating an interaction induced elimination of Fermi pockets. Before
${\rm NdNiO_2}$ enters correlated insulator, it is a compensated metal with one small electron pocket formed by $d_{xy}$
orbital of Nd and four small hole pockets formed by $d_{z^2}$ orbital of Ni. The estimated phase transition temperature (${T_N}$)
from PM phase to $(\pi, \pi, \pi)$ AFM phase is 70 $\sim$ 90 K for moderate interaction strength of U = 5 $\sim$ 6 eV.

Through these studies, we identify two key messages that are distinguishable from the cuprates:
1) ${\rm NdNiO_2}$ is dominated by the physics of Mott-Hubbard instead of charge-transfer;
2) effective exchange coupling parameters are about one-order smaller than those of cuprates.
In this regarding, supposed that the ground state is magnetic, our calculations demonstrate $(\pi,\pi,\pi)$ AFM ordering is energetically favorable.
Moreover, our results provide a natural understanding of two experimental observations.
First, $3d$-electrons of Ni tend to form AFM ordering around 70 $\sim$ 90 K, coinciding with the minimum in resistivity and the drop in Hall coefficient.
Second, the $(\pi,\pi,\pi)$ AFM ordering could be weak (compared with cuprates), because of the small
effective exchange coupling and the hybridization with itinerant 5$d$-electrons of Nd.
This could be the reason why AFM ordering is missing in previous study, which calls for more
careful neutron scattering measurements on ${\rm NdNiO_2}$.

\begin{figure*}
\includegraphics[width=14cm]{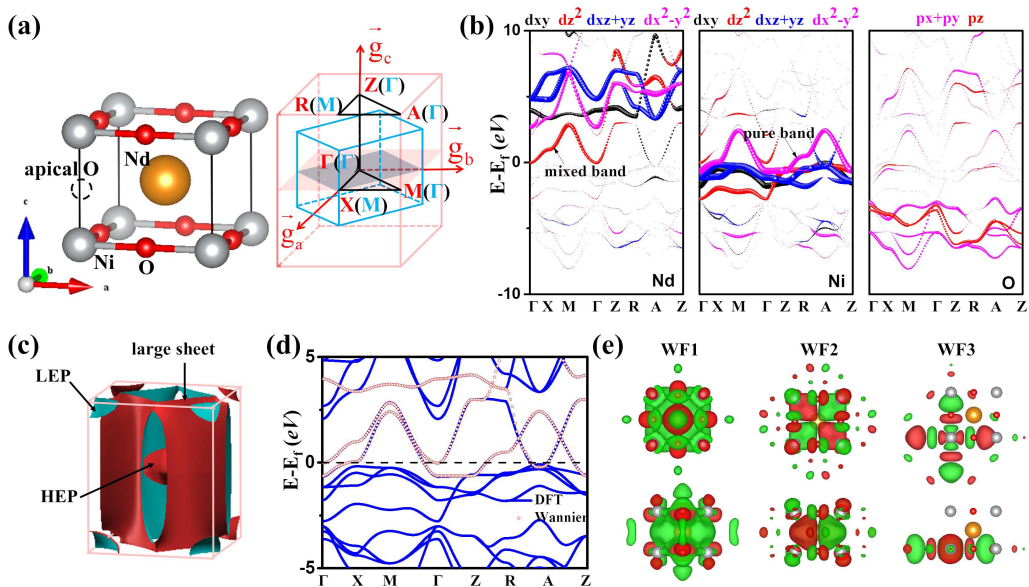}
\caption{(a) Atomic structure of tetragonal ${\rm NdNiO_2}$ and first Brillouin zone. The red and blue lines (labels) denotes the
first Brillouin zone of PM phase and $(\pi, \pi, \pi)$ AFM phase, respectively. The high symmetric line for band calculation are $\Gamma$(0, 0, 0)-X(0.5, 0, 0)-M(0.5, 0.5, 0)-$\Gamma$-Z(0, 0, 0.5)-R(0.5, 0, 0.5)-A(0.5, 0.5, 0.5)-Z. (b) Orbital resolved band structure of PM phase. The $d_{xy}$, $d_{z^2}$, $d_{xz+yz}$ and $d_{x^2-y^2}$ of Nd (spin-up Ni, spin-down Ni) are marked by black, red, blue and pick filled circles. The $p_{z}$ and $p_{x+y}$ of O are marked by red, blue and pick filled circles. The size of circles represents the orbital weights.
(c) Perspective view of Fermi surfaces. LEP and HEP denotes light and heavy electron pocket, respectively. (d) The comparison between first-principles and
Wannier-fitting bands around the Fermi-level. (e) Top and side view of three maximally localized Wannier functions.}
\end{figure*}

%\section{II. Methods}

The first-principle calculations are carried out with the plane wave projector augmented wave method as implemented in the Vienna \textit{ab initio} simulations
package (VASP) \cite{vas1,vas2,vas3}. The Perdew-Burke-Ernzerhof (PBE) functionals of generalized gradient approximation (GGA) is used for PM phase \cite{vas4}.
To incorporate the electron-electron interactions, DFT + U is used for AFM phase, which can reproduce correctly the gross features of correlated-electrons in
transition metal oxides \cite{ldau1, ldau2, ldau3}. The 4$f$ electrons of ${\rm Nd^{3+}}$ are expected to display the local magnetic moment as ${\rm Nd^{3+}}$ in ${\rm Nd_{2}CuO_{4}}$ \cite{Ndf} and are treated as the core-level electrons. The Hubbard U (0 $\sim$ 8 eV) term
is added to 3$d$ electrons of Ni. The energy cutoff of 600 eV, and Monkhorst-Pack $k$ point mesh of ${11\times11\times11}$ and ${18\times18\times30}$ is used for PM and
AFM phase, respectively. The maximally localized Wannier functions (WFs) are constructed by using Wannier90 package \cite{wan1, wan2}. The structure of infinite-layer
${\rm NdNiO_2}$ is shown in Fig. 1(a), including ${\rm NiO_2}$ layers sandwiched by ${\rm Nd}$, which can be obtained from the perovskite ${\rm NdNiO_3}$
with reduction of apical O atoms in $c$ direction \cite{exp1, exp2}. Due to apical O vacancies, the lattice constant in $c$ direction shrinks (smaller than $a$ direction)
and the space group becomes $P4/mmm$. The experimental lattice constant $a$ = $b$ = 3.92 {\AA} and $c$ = 3.28 {\AA} are used in our calculations.

%\section{III. Results}

%\subsection{A. Electronic structure in PM phase}
Firstly, We present the band structure of PM phase without Hubbard U.
The orbital resolved band structure of PM phase is shown in Fig. 1(b). Comparing with typical cuprates ${\rm CaCuO_{2}}$ \cite{CaCuO}, two significant differences are noted:
1) there is a gap $\sim$ 2.5 eV between $2p$ orbitals of O and $3d$ orbitals of Ni. According to Zaanen-Sawatzky-Allen classification scheme \cite{ZSA}, this indicates that
the physics of ${\rm NdNiO_2}$ is close to Mott-Hubbard rather than charge-transfer; 2) there are two bands crossing the Fermi level, in which
one is mainly contributed by $d_{x^2-y^2}$ orbital of Ni (called pure-band) and the other one has a complicated orbital compositions (called mixed-band).
In ${k_z}$ = 0 plane, the mixed-band is mainly contributed by $d_{z^2}$ orbital of Nd and Ni. The dispersion around $\Gamma$ point is relatively small,
called heavy electron pocket (HEP). In ${k_z}$ = 0.5 plane, the mixed-band is mainly contributed by $d_{xy}$ ($d_{xz}$, $d_{yz}$ and $d_{z^2}$) orbital of Nd (Ni).
The dispersion around A point is relatively large, called light electron pocket (LEP). As a comparison, one notices that there is only one pure-band crossing the Fermi level
in ${\rm CaCuO_{2}}$ \cite{CaCuO}. The Fermi surface of PM phase is shown in Fig. 1(c). There is a large sheet contributed by the pure-band, as the case in ${\rm CaCuO_{2}}$ \cite{CaCuO}.
This Fermi surface is obviously two-dimensional (2D), because of the weak dispersion along $\Gamma$-Z. In addition, there are two electron pockets residing
at $\Gamma$ and A point, respectively, showing a feature of three-dimensional (3D) rather than 2D (see labels HEP and LEP in Fig. 1(c)).
Therefore, the 3D metallic state will be hybridized with the 2D correlated state in ${\rm NiO_{2}}$ plane, suggesting ${\rm NdNiO_{2}}$ to be an "oxide-intermetalic" compound \cite{om1, om2}.

%\begin{SCfigure*}
\begin{figure*}
\includegraphics[width=14cm]{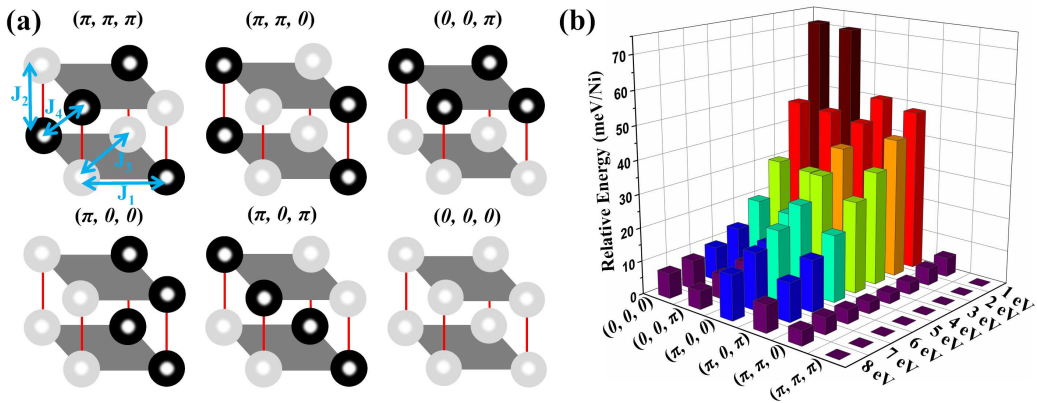}
\caption{(a) Illustration of six collinear spin configurations. The while and black ball represents local up and down spin moment, respectively.
The four exchange coupling parameters are indicated by the blue arrows. (b) Energy comparison for the six collinear spin configurations with different values of Hubbard U. Energy of $(\pi, \pi, \pi)$ AFM is set to zero.}
\end{figure*}
%\end{SCfigure*}

The existence of mixed-band also reflects the inherent interactions between Nd $5d$ and Ni $3d$ electrons. To explore the low energy physics of ${\rm NdNiO_2}$,
a three-band model consisting of Ni $d_{x^2-y^2}$, Nd $d_{z^2}$ and Nd $d_{xy}$ orbitals is constructed by Wannier90 package. As shown Fig. 1(d), one can see the good agreement between first-principles and Wannier-fitting bands near
the Fermi level. The corresponding three maximally localized WFs are shown in Fig. 1(e), demonstrating the main feature of $d_{z^2}$ (WF1) and $d_{xy}$ (WF2) orbital
of Nd, and $d_{x^2-y^2}$ (WF3) orbital of Ni. However, these WFs still have some derivations from standard atomic orbitals, that is, WF1 and WF2 are mixed with
$d_{z^2}$ orbital of Ni, and WF3 is mixed with ${p_{x/y}}$ orbital of O in the ${\rm NiO_{2}}$ plane. According to the classical Goodenough-Kanamori-Anderson rules \cite{GKA1, GKA2, GKA3},
these derivations (or hybridizations) will give clues for the magnetic properties.

%\subsection{B. Magnetic properties}
To determine the magnetic ground state of ${\rm NdNiO_2}$, six collinear spin configurations are taken into account in a
${2 \times 2 \times 2}$ supercell, that is, AFM1 with \textbf{q = $(\pi, \pi, \pi)$}, AFM2 with \textbf{q = $(\pi, \pi, 0)$}, AFM3 with \textbf{q = $(0, 0, \pi)$}, AFM4 with \textbf{q = $(\pi, 0, 0)$},
AFM5 with \textbf{q = $(\pi, 0, \pi)$} and FM with \textbf{q} = (0, 0, 0), as shown in Fig. 2(a). Within all Hubbard U ranges, we found that AFM1 configuration always has the lowest energy, as shown in Fig. 2(b), indicating a stable $(\pi, \pi, \pi)$ AFM phase with respect to electron-electron interactions and is in accordance with random phase approximation treatment \cite{RPA}. This can be attributed to the special orbital distributions around the Fermi level. The intralayer NN
exchange coupling is the typical 180$^{\circ}$ typed Ni-O-Ni superexchange coupling, that is, the coupling between $d_{x^2-y^2}$ orbital of Ni is mediated by ${p_{x/y}}$ orbital of O (see WF3), preferring a $(\pi, \pi)$ AFM phase in ${\rm NiO_2}$ plane. The interlayer NN exchange coupling 
is due to the superexchange between the Ni $d_{z^2}$ orbitals mediated by Nd $d_{z^2}$ orbital as shown in WF1, preferring a $(\pi, \pi)$ AFM phase between ${\rm NiO_2}$ planes. Therefore, the superexchange coupling
results in a stable $(\pi, \pi, \pi)$ AFM phase in ${\rm NdNiO_2}$. Moreover, the magnetic anisotropy is further checked by including the spin-orbit coupling (SOC). We found that the spin moment
prefers along $c$ direction with the magnetic anisotropic energy of $\sim$0.5 meV/Ni. Thus, the tiny SOC effect can be safely neglected in the following phase transition temperature calculations.

%\begin{SCfigure*}
\begin{figure*}
\includegraphics[width=14cm]{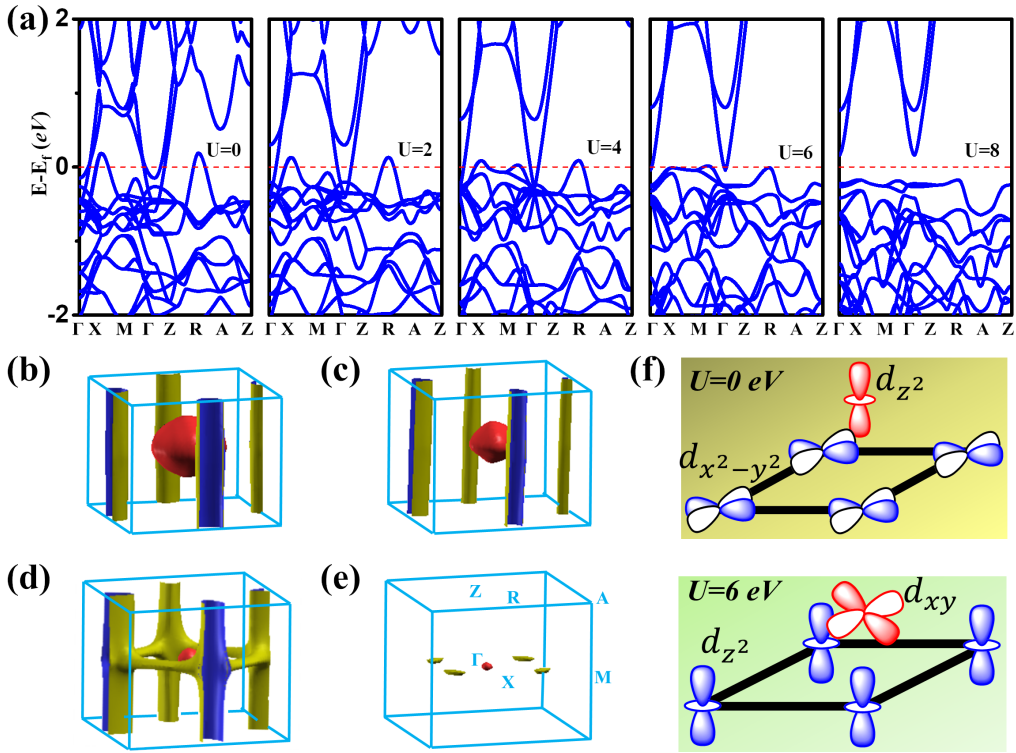}
\caption{(a) Band structures of $(\pi, \pi, \pi)$ AFM phase with different values of Hubbard U. (b)-(e) Perspective view of the Fermi surfaces of $(\pi, \pi, \pi)$ AFM with U = 0, 2, 4 6 eV. The two hole pockets in (b)-(d) are degenerated and can not be distinguished from this picture. The high symmetry k-points in (e) are labelled to guide the eye. (f) Schematic diagram of major self-doping channel at U = 0 and U = 6 eV. The red/blue color represents $d$ orbital of Nd/Ni respectively.}
\end{figure*}
%\end{SCfigure*}

In cuprates, the Fermi surface is unstable with electron-electron interactions, making its parent phase to be an AFM insulator. However, this
is apparently not the case in ${\rm NdNiO_2}$, because of the extra electron pockets and the inherent interaction between Nd $5d$ and Ni $3d$ electrons. At U=0 eV, there are two electron pockets at $\Gamma$ point and two hole pockets along X-R direction as shown in Fig. 3(a)-(b). Physically, the origin of these four pockets can be easily understood through the comparison of orbital resolved band structures between PM phase (Fig. 1(b)) and
$(\pi, \pi, \pi)$ AFM phase (Fig. 4). Because of the Zeeman field on Ni, its spin-up and -down bands are split away from each other. The original
pure-band ($d_{x^2-y^2}$ orbital of Ni) in PM phase becomes partially occupied in spin-up channel (forming two hole pockets) and totally unoccupied
in spin-down channel. Hence, the two hole pockets in AFM phase are inherited from large sheet in PM phase, showing a 2D character with neglectable dispersion
along $\Gamma$-Z direction. For the electron pockets at $\Gamma$ point, the heavier one is mainly contributed by $d_{z^2}$ orbital of Nd and Ni, so it comes
from the HEP at $\Gamma$ point of PM phase. While for the lighter one, it comes from the LEP at A point of PM phase which is folded into the $\Gamma$
point of $(\pi, \pi, \pi)$ AFM phase [see Fig. 1(a)]. The orbital composition can also be used to check this folded band, which is contributed by
$d_{xy}$ orbital of Nd, $d_{xz/yz}$ ($d_{z^{2}}$) orbital of spin-down (-up) Ni.

These pockets have a different evolution with the increasing value of Hubbard U. For electron pockets, the heavier one is very sensitive to Hubbard U and disappears at U = 1 eV. Meanwhile the lighter one doesn't appear until U = 6 eV. In addition, the orbital components of lighter electron pockets are purified by electron-electron interaction and it mainly contributed by $d_{xy}$ of Nd in the large U limit as shown in Fig. 4.
The case for hole pockets is rather complicated. Firstly, the bands of hole pockets become flat with the
increasing value of Hubbard U. Secondly, the original hole pockets formed by $d_{x^2-y^2}$ orbital of Ni gradually disappear, meanwhile, a new hole
pocket formed by $d_{z^2}$ orbital of Ni appears along $\Gamma$-M as shown in Fig. 3(d). At U = 6 eV, ${\rm NdNiO_2}$ is a compensated
metal with a small electron pocket at $\Gamma$ point and four hole pockets along $\Gamma$-M as displayed in Fig. 3(e). Further increasing the
value of Hubbard U, the system enters an AFM insulator, just like cuprates. Therefore, the metal-to-insulator phase
transition point is near U $\sim 6$ eV. If Hubbard U is less than 6 eV, ${\rm NdNiO_2}$ is an AFM metal with relatively small amount of holes that are
self-doped \cite{RPA, sd1, sd2, sd3, om1} into $d$ orbitals of Ni. Interestingly, there is an orbital shift from $3d_{x^2-y^2}$ orbital of Ni at U = 0 eV to $3d_{z^2}$ orbital of Ni at U = 6 eV in the ${\rm NiO_2}$ plane, as depicted in Fig. 3(f). We speculate that this orbital shift may change the paradigm after doping \cite{pre2, dz2}. 
Moreover, without the Hubbard U, the $2p$ orbital of O is far away from the Fermi level, just like the case of PM phase.
However, with the increasing value of Hubbard U, the gap between $3d$ orbital of Ni and $2p$ orbital of O gradually decreases (Fig. 4),
demonstrating an evolution from Mott-Hubbard metal to charge-transfer insulator.

%\begin{SCfigure*}
\begin{figure*}
\includegraphics[width=14cm]{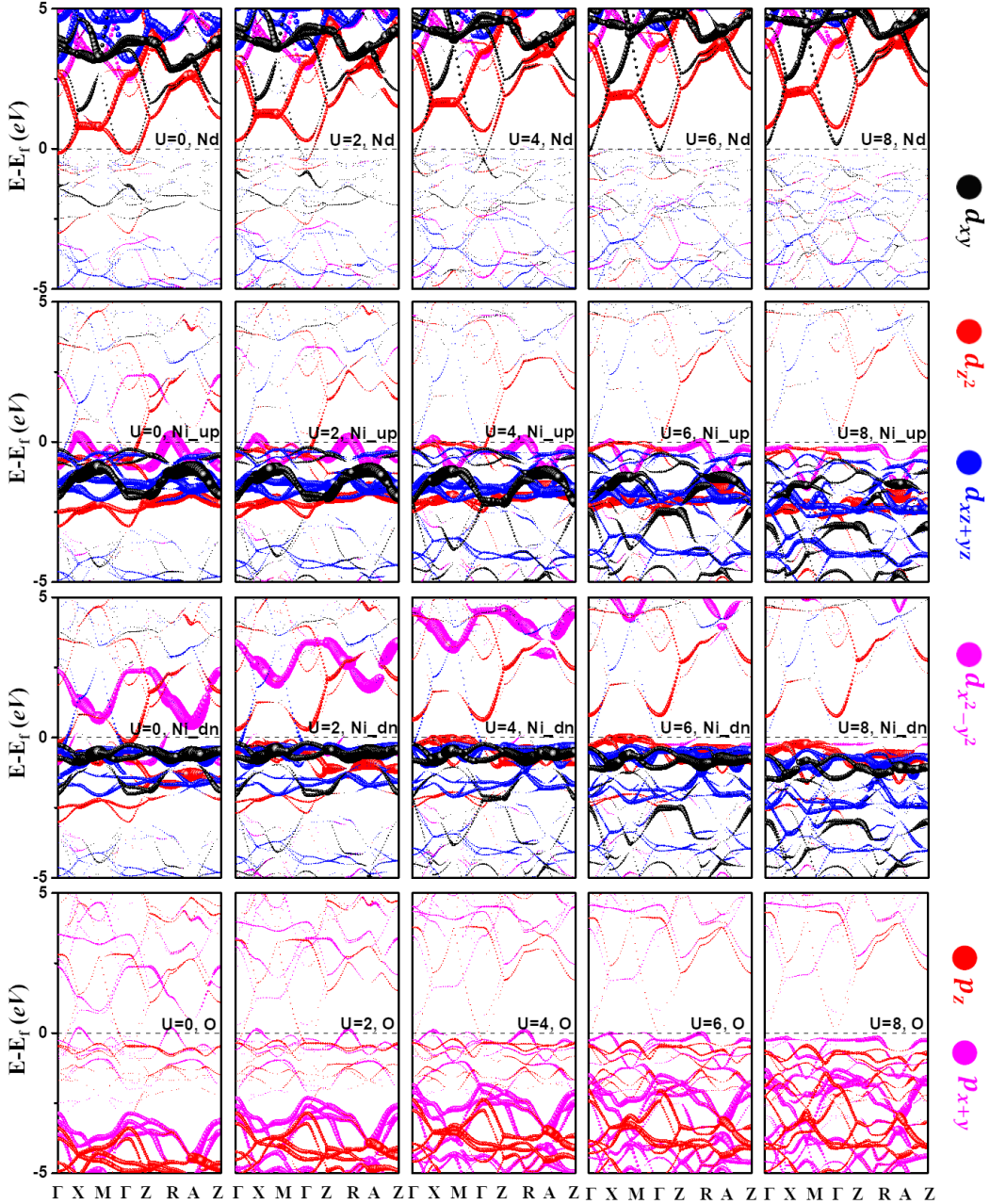}
\caption{ Orbital resolved band structures of $(\pi, \pi, \pi)$ AFM phase with different values of Hubbard-U.
The first, second, third and forth row represents Nd, spin-up Ni, spin-down Ni and O, respectively. The filled circles with different colors have the same meaning as those in Fig. 1. The name of $d$ orbitals in the AFM supercell has been aligned to that of unit cell.}
\end{figure*}
%\end{SCfigure*}

%\subsection{C. PM-AFM transition}

In order to quantitatively describe such a phenomenon, the phase transition temperature is further calculated. For $(\pi, \pi, \pi)$ AFM phase,
the magnetic momentum of ${\rm Ni}$ increases from 0.58 ${\mu_B}$ (U = 0 eV) to 1.04 ${\mu_B}$ (U = 8 eV) and becomes gradually
saturated, as shown in Fig. 5(a). This is also consistent with the fact that $d_{x^2-y^2}$ orbital of Ni is closer to
single occupation with the increasing value of Hubbard U. Therefore, ${\rm Ni}$ is spin one half (${S = 1/2}$)
in infinite-layer ${\rm NdNiO_2}$, just like the case in cuprates. To extract the exchange coupling parameters of
${J_1}$, ${J_2}$, ${J_3}$ and ${J_4}$ (as labelled in Fig. 2(a)), the total energy of five AFM configurations
obtained from DFT + U calculations are mapped onto the Heisenberg spin Hamiltonian. In the ${2 \times 2 \times 2}$ supercell, there are 8 Ni atoms and the total energy of different AFM configurations are:

{\begin{small}
\begin{equation}
\begin{aligned}
{E_{\rm AFM1}}=&{E_0}-16{J_1}{S^2}-8{J_2}{S^2}+16{J_3}{S^2}+32{J_4}{S^2}\\
{E_{\rm AFM2}}=&{E_0}-16{J_1}{S^2}+8{J_2}{S^2}+16{J_3}{S^2}-32{J_4}{S^2}\\
{E_{\rm AFM3}}=&{E_0}+16{J_1}{S^2}-8{J_2}{S^2}+16{J_3}{S^2}-32{J_4}{S^2}\\
{E_{\rm AFM4}}=&{E_0}+8{J_2}{S^2}-16{J_3}{S^2}+32{J_4}{S^2}\\
{E_{\rm AFM5}}=&{E_0}-8{J_2}{S^2}-16{J_3}{S^2}-32{J_4}{S^2}
\end{aligned}
\end{equation}
\end{small}}

\noindent where ${E_0}$ is the reference energy without magnetic order.
The calculated exchange coupling parameters as a function of Hubbard U are shown in Fig. 5(b). We would like to make several remarks here:
1) the NN intralayer exchange coupling (${J_1}$), mediated by $d_{x^2-y^2}$ orbital of Ni, demonstrates a $1/U$ law;
2) the NN interlayer exchange coupling (${J_2}$) is $\sim$ 10 meV with little variation. The positive value of
${J_2}$ indicates an AFM coupling between ${\rm NiO_2}$ planes; 
3) the next NN intralayer exchange coupling (${J_3}$), mediated by $d_{xy}$ orbital of Nd, is comparable to ${J_1}$ at large value of Hubbard U, which is dramatically different to that in infinite-layer ${\rm SrFeO_2}$ \cite{SFO1, SFO2, SFO3}. This large value could be attributed to the relative robustness of lighter electron pocket and orbital purification; 
4) the next NN interlayer exchange coupling (${J_4}$) is $\sim$ 0 meV, indicating the validity of our Hamiltonian up to the third NN;
5) ${J_1}$, ${J_2}$ and ${J_3}$ have the same strength at large U, suggesting ${\rm NdNiO_2}$ is a 3D magnet rather than 2D magnet.

%\begin{SCfigure*}
\begin{figure*}
\includegraphics[width=14cm]{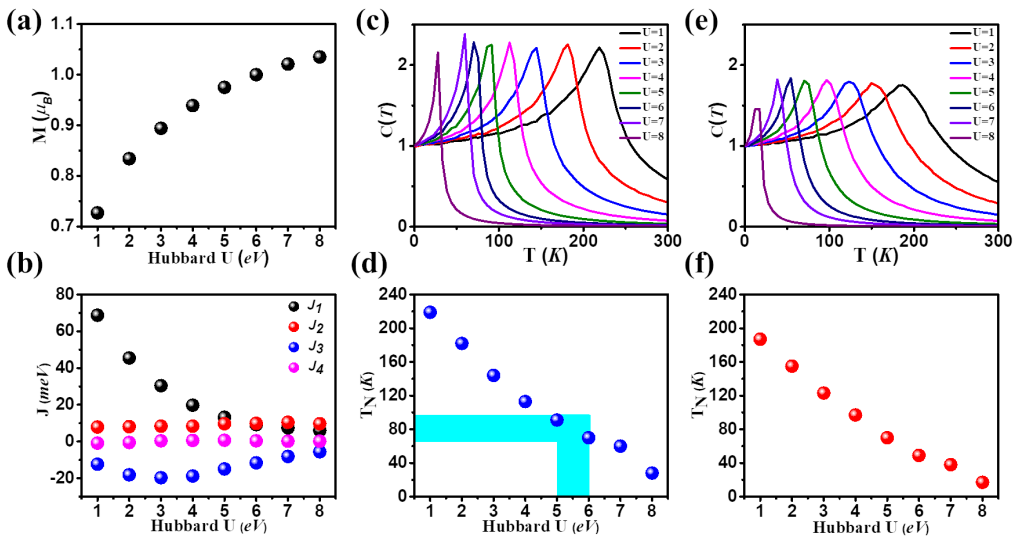}
\caption{The Hubbard U dependence of (a) the magnetic momentum of ${\rm Ni}$ in $(\pi, \pi, \pi)$ AFM phase,
(b) the exchange coupling parameters, (c) specific heat ($C$) vs $T$ with four J's, (d) the estimated ${T_N}$ with four J's, (e) specific heat ($C$) vs $T$ with ${J_1}$, ${J_3}$ only and (f) the estimated ${T_N}$ with ${J_1}$, ${J_3}$ only.
The shaded region in (d) highlights the possible ${T_N}$ of
70 $\sim$ 90 K with a reasonable U = 5 $\sim$ 6 eV.}
\end{figure*}
%\end{SCfigure*}

Based on the above exchange coupling parameters, the phase transition temperature (${T_N}$) is calculated by classical Monte Carlo method in a ${12 \times 12 \times 12}$ supercell based on the classical spin Hamiltonian:

\begin{small}
\begin{equation}
H=\sum_{\langle i,j\rangle}J_{ij}\vec{S_i}\cdot\vec{S_j}
\end{equation}
\end{small}

\noindent where the spin exchange parameters $J_{ij}$ have been defined above.
First, we calculate the specific heat (\textit{C}) after
the system reaches equilibrium at a each given temperature ($T$), as shown in Fig. 5(c). Then, ${T_N}$ is extracted from the peak position in the
curve of $C(T)$, as shown in Fig. 5(c). For U = 1 eV, ${T_N}$ is as high as 220 K, which can be ascribed to the
large value of ${J_1}$. With the increasing value of Hubbard U, ${T_N}$ gradually decreases and becomes $\sim$ 70 K at U = 6 eV. 
To further check the effect of interlayer exchange coupling on 3D magnet, an additional Monte Carlo
calculation is performed without ${J_2}$ and ${J_4}$. As shown in Fig. 5(e), the $C(T)$ vs $T$ plot shows a broaden peak at a lower temperature. Since Mermin-Wagner theorem prohibit magnetic order in 2D isotropic Heisenberg model at any nonzero temperatures \cite{MWt}, 
the broad peak in $C(T)$ vs $T$ plot implies the presence of short-range order. Regarding the small drop of ${T_N}$ (about 30 K in Fig. 5(f)), our MC simulations indicate that the weak interactions between ${\rm NiO_2}$ planes.

%\section{Summary and Discussion}
Although the exact value of Hubbard U cannot be directly extracted from the first principles calculations,
its value range can still be estimated based on similar compounds. The infinite-layer nickelates are
undoubtedly worse metals compared to elemental nickel with U $\sim$ 3 eV \cite{UNi}, which can be considered as
a lower bound of Hubbard U. The Coulomb interaction in infinite-layer nickelates should be smaller than that
in the charge-transfer insulator NiO with U $\sim$ 8 eV \cite{ldau1}, which can be considered as a upper bound of Hubbard U.
Therefore, a reasonable value of Hubbard U in ${\rm NdNiO_2}$ will between 3 eV and 8 eV.
In the following discussions, we use U = 5 $\sim$ 6 eV \cite{ldau4, om1} to draw our conclusions:
1) with the decreasing of temperature, there is a phase transition from PM phase to $(\pi, \pi, \pi)$ AFM phase near 70 $\sim$ 90 K;
2) the exchange-coupling parameters are $\sim$ 10 meV, which is one order smaller than cuprates \cite{J1, J2, J3, J4, J5} and results in a low ${T_N}$ compared with cuprates;
3) the self-doing effect from ${5d}$ orbital of Nd and ${3d}$ orbital of Ni may screen the local magnetic momentum in $d_{x^2-y^2}$ orbital of Ni, which gives a small
magnetic momentum less than 1 ${\mu_B}$ and makes the long-range AFM order unstable \cite{exp1, exp2, Nat1, Mei1};
4) the Fermi surface of PM phase is quite large with two 3D-liked electron pockets, while the Fermi surface of $(\pi, \pi, \pi)$ AFM phase is quite small
with one 3D-liked electron pocket and four 2D-liked hole pocket.  Therefore, %accompanied by the phase transition, 
there could exist a crossover from normal metal to bad AFM metal around $T_N\sim 70 - 90$ K, 
which provides a plausible understanding of minimum of resistivity and Hall coefficient drop in infinite-layer ${\rm NdNiO_2}$\cite{Nat1}.
We envision that our calculations will intrigue intensive interests for studying the magnetic properties of high quality infinite-layer ${\rm NdNiO_2}$ samples.

Lastly, we would like to make some remarks on the existing experiments.
Some recent experiments fail to find bulk superconductivity in
${\rm NdNiO_2}$ systems, and the parent samples show strong insulating behaviors \cite{Wen1}.
The insulating behavior could be attributed to strong inhomogenious disorder
or improper introduction of H during the reaction with CaH$_2$ \cite{Hins}.
Especially, it is worth noting, the experiments cannot rule out the possibility of weak AFM ordering,
due to the presence of Ni impurities in their samples.
(Actually this problem has been pointed out before \cite{exp1, exp2}.)
The strong ferromagnetic order from elemental Ni
would dominate over and wash out the weak signal of AFM ordering from ${\rm NdNiO_2}$ as we suggested in this work.
In this regarding, the upcoming inelastic neutron scattering on high-quality samples is highly desired.

%\textit{Acknowledgement.---}
W.Z. thanks Chao Cao for sharing their unpublished DMFT results,
and thanks Filip Ronning, H. H. Wen, G. M. Zhang for helpful discussion.
This work was supported by NSFC (No. 11774325, 21603210, 21603205, 21688102),
National Key Research and Development Program of China (No. 2017YFA0204904, 2016YFA0200604),
Anhui Initiative in Quantum Information Technologies (No. AHY090400),
Fundamental Research Funds for the Central Universities and the Start-up Funding
from Westlake University. We thank Supercomputing Center at USTC for providing
the computing resources.

%\newpage{}%

\end{document}